\documentclass[aps,prb, twocolumn,superscriptaddress]{revtex4-2}

\usepackage[utf8]{inputenc}
\usepackage{amsthm}
\usepackage{amsmath}
\usepackage{xfrac}
\usepackage{color}
\usepackage{soul,xcolor}
\usepackage{hyperref}
\usepackage{graphicx,xcolor}
\usepackage{amssymb}
\usepackage{bbold}
\usepackage{gensymb}
\usepackage{dsfont}
\usepackage{float}
\usepackage{pifont}

\setstcolor{red}

\usepackage[normalem]{ulem}
\newcommand{\commentout}[1]{}

\newcommand{\ii}{\mathrm{i}} 
\newcommand{\eul}{\mathrm{e}} 
\newcommand{\id}{\mathbb{1}} 

\newcommand{\ketbra}[2]{|#1\rangle\!\langle #2|} 
\newcommand{\tr}{\mathrm{tr}} 

\definecolor{darkgreen}{RGB}{0,128,0}

\begin{document}

\title{Benchmarking Floquet Master Equations for Periodically Driven Open Quantum Systems}

\author{Konrad Mickiewicz}
\affiliation{Institut f{\"u}r Theoretische Physik, Technische Universit{\"a}t Dresden, 
D-01062, Dresden, Germany}

\author{Valentin Link}
\affiliation{Institut für Physik und Astronomie, Technische Universität Berlin, D-10623, Berlin, Germany}

\author{Walter T. Strunz}
\affiliation{Institut f{\"u}r Theoretische Physik, Technische Universit{\"a}t Dresden, 
D-01062, Dresden, Germany}

\begin{abstract}
The dynamics of open quantum systems is commonly described by quantum master equations derived under the assumption of weak system-bath coupling and a separation of timescales between system and bath. When the system is additionally subjected to a periodic driving, the validity of the resulting Floquet master equations is further restricted to regimes of weak or high-frequency driving. Here, we benchmark a set of commonly used Floquet master equations for a model of two locally driven spins coupled to a shared Ohmic reservoir at finite temperature. We systematically probe the accuracy of the equations as a function of the driving parameters, thus identifying limits of their applicability. Dynamical maps predicted by each master equation are compared against numerically exact non-Markovian simulations, tracking the full relaxation dynamics. We find that the accuracy of each master equation closely reflects the assumptions underlying its derivation. For the Floquet-Lindblad equation, errors can be strongly amplified near resonances where the secular approximation breaks down, while approaches that avoid the secular approximation perform better and exhibit a more systematic dependence of the error on driving frequency and amplitude. 
\end{abstract}\maketitle

\section{Introduction}

Periodic driving is a a central tool for realizing non-equilibrium dynamics in physical systems. An external drive explicitly breaks detailed balance and can thus be used to steer a system into states that are inaccessible in equilibrium \cite{oka_floquet_2019, rudner_floquet_2020} or to engineer exotic interactions in many-body systems \cite{eckardt_colloquium_2017}. In quantum systems, specific examples include ultracold gases in optical lattices \cite{weitenbergTailoringQuantumGases2021}, cavity QED systems \cite{longBoostingQuantumState2022,zhou_cavity_2024}, and superconducting circuits \cite{gandon_engineering_2022, fazioManybodyOpenQuantum2025}. In practice, energy injected by the drive is absorbed by the surrounding environment, thereby stabilizing a time-periodic Floquet steady state at long times \cite{Shirai2016May,engelhardt2019Discontinuities,Sato2020Oct, petiziolCavityBasedReservoirEngineering2022a, li2024Coherent, Ritter2025}. Since these states are stabilized autonomously through the interplay of drive and dissipation, they are of particular interest for quantum control protocols \cite{Bai2021control}, quantum thermodynamics \cite{koyanagi2022Numerically,boettcher2024Dynamics,kolisnyk_floquet_2024,Alamo2024Jun, alamoMinimalQuantumHeat2024a, boettcherDynamicsStronglyCoupled2024}, and quantum state engineering \cite{wanckelDissipativeFloquetEngineering2026}. 

An accurate description of dissipation in the presence of driving is generally nontrivial and requires a combination of methods from quantum Floquet theory \cite{bukov_universal_2015} and open quantum systems \cite{vacchiniOpenQuantumSystems2024}. Floquet master equations \cite{Ketzmerick2010Aug, schnellHighfrequencyExpansionsTimeperiodic2021a, mori_floquet_2023} are a common approach to this problem. They govern the evolution of the reduced system state obtained by tracing out the reservoir entirely. Such a compact representation is conceptually appealing, but the derivation of master equations relies on strong assumptions on the structure of the dissipation, most prominently weak system-bath coupling and, in the case of Lindblad master equations, the secular approximation. Beyond these, Floquet master equations usually require additional approximations with respect to the time-dependent part of the Hamiltonian \cite{mori_floquet_2023}, restricting their validity to regimes of weak or high-frequency driving. The goal of this paper is to investigate how the accuracy of various Floquet master equations depends on the driving parameters.

In contrast to perturbative master equations, numerically exact methods for non-Markovian open quantum systems can capture the full relaxation dynamics of driven open systems without additional assumptions on the drive \cite{mickiewicz_exact_2025}. These exact methods represent the dissipation through a small set of auxiliary degrees of freedom engineered to reproduce the full effect of the bath on the system dynamics, and allow for the inclusion of a local drive without additional overhead. Here, we employ the uniform time-evolving matrix product operator (uniTEMPO) method \cite{linkOpenQuantumSystem2024a, kahlertSimulatingLandauZener2024, mickiewicz_exact_2025}, an improved variant of the TEMPO approach \cite{strathearnEfficientNonMarkovianQuantum2018, Jorgensen2019Dec}, in order to compute an exact reference dynamics, thus allowing us to determine the exact errors of the various master equations. 

While exact methods can be highly efficient for the smaller systems, like the one considered in this paper, their numerical cost scales unfavorably with the system size \cite{keelingProcessTensorApproaches2025}. In the weakly coupled regime, master equations therefore remain an important theoretical and numerical tool, especially in many-body systems, as they avoid the inclusion of auxiliary degrees of freedom. Moreover, master equations can provide a more transparent physical intuition for dissipative processes through explicit analytical expressions for decay rates and jump operators. 

In this work, we benchmark the accuracy of various Floquet master equations for a non-integrable example problem, a periodically driven two-spin-boson model. Accuracy assessments of undriven master equations with respect to coupling strength and bath structure have been carried out previously, for example in Ref.~\cite{hartmann_accuracy_2020}. Here, we focus specifically on Floquet master equations and the dependence of their accuracy on the driving parameters, in particular the driving amplitude and driving frequency. We determine how the additional approximations required for a periodic driving affect the accuracy of each approach. These findings provide guidance for choosing the appropriate master equation in a given parameter regime.

The paper is structured as follows. In Sec.~\ref{sec:two-spin-boson} we introduce the periodically driven two-spin-boson model. In Sec.~\ref{sec:master_equations}, we present derivations of the relevant Floquet master equations, highlighting the approximations involved. In Sec.~\ref{sec:error_convergence} we introduce a rigorous error measure and demonstrate convergence of the reference solution. Finally, in Sec.~\ref{sec:results} we present the main results, organized into three parameter regimes: weak driving, strong driving, and low temperature.

\section{Two-spin-boson model} \label{sec:two-spin-boson}

In this work we consider open quantum systems with local driving, i.e.~where the time-dependence only acts locally on the system part of the Hamiltonian and not on the reservoir. In particular, the total Hamiltonian can be written as
\begin{equation}\label{eq:H_total}
    H_\mathrm{tot}(t) = H_{\mathrm{sys}}(t)\otimes \id_\mathrm{env} + H_\mathrm{int} + \id_\mathrm{sys}\otimes H_\mathrm{env}.
\end{equation}
We assume a Gaussian bosonic bath with linear coupling which can be described by a collection of bosonic modes
\begin{equation}
    H_\mathrm{env}=\sum_\lambda \omega_\lambda b_\lambda^\dagger b_\lambda
\end{equation}
\begin{equation}\label{eq:H_total}
    H_\mathrm{int} =  L\otimes\sum_\lambda g_\lambda (b_\lambda^\dagger+ b_\lambda) 
\end{equation}
where $b_\lambda^{(\dagger)}$ are bosonic annihilation (creation) operators, $g_\lambda$ denotes the coupling strength to the bath mode $\lambda$ and $L = L^\dagger$ is the system coupling operator. The system Hamiltonian includes a time-independent part ($H_0$) and a drive
\begin{equation}
\begin{split}
    H_{\mathrm{sys}}(t) &= H_0 + H_\mathrm{drive}(t).
\end{split}
\end{equation}
We assume the drive to be periodic with frequency $\omega_d=2\pi/T$ such that a Fourier series expansion exists
\begin{equation}
    H_\mathrm{drive}(t)=H_\mathrm{drive}(t+T)=\sum_{l\neq0}H_l\eul^{\ii l\omega_d t}.
\end{equation}
As a concrete example, we will consider a model of two non-interacting spins $A$ and $B$ coupled to a common Gaussian bosonic environment, a model that has also been used in Ref.~\cite{hartmann_accuracy_2020} for benchmarking master equations in the case of no driving. The time-independent part of the Hamiltonian reads 
\begin{equation}
    H_0=\frac{\omega_A}{2}\sigma^A_x + \frac{\omega_B}{2}\sigma^B_x
\end{equation}
with spin frequencies $\omega_A$ and $\omega_B$. We assume a transverse drive with driving amplitude $\epsilon_d$ and the driving frequency $\omega_d$ acting identically on both spins
\begin{equation}
    H_\mathrm{drive}(t)= \frac{\epsilon_d}{2}\cos(\omega_dt)(\sigma^A_z+\sigma^B_z).
\end{equation}
The spins couple to the bath transversely via
\begin{equation}
    L= \frac{1}{2}(\sigma_z^A+\sigma_z^B).
\end{equation}
The choice of the two-spin-boson model allows us to study the influence of the detuning of the spins on the validity of various master equation approaches. Generally, to derive a master equation of Gorini–Kossakowski–Sudarshan–Lindblad (GKSL) form from a Redfield master equation, one needs to perform the so-called secular approximation. In the resonant case, i.e. $\omega_A=\omega_B$, the singlet and triplet subspaces are degenerate, resulting in overall better justification of the secular approximation. In the case of a finite detuning, i.e. $\omega_A \neq \omega_B$, this degeneracy is lifted and $H_0$ has more transition frequencies, possibly making the secular approximation less accurate. 

The Gaussian bath is characterized by the spectral density $J(\omega) = \sum_\lambda |g_\lambda|^2\delta(\omega-\omega_\lambda)$, which defines the bath correlation function with an inverse temperature $\beta$
\begin{equation}
\begin{split}
    &\alpha(t) = \int_0^\infty \mathrm{d}\omega J(\omega) \big(\coth(\beta\omega/2)\cos(\omega t)-\ii \sin(\omega t) \big).
\end{split}
\end{equation}
In this work we consider an Ohmic spectral density with exponential cutoff 
\begin{equation}
    J(\omega) = \alpha \omega \mathrm{e}^{-\omega/\omega_c},
\end{equation}
where $\alpha$ is a dimensionless coupling-strength and $\omega_c$ is a high-frequency cutoff. It is convenient to introduce the one-sided Fourier transform of the bath correlation function
\begin{equation} \label{eq:F}
    \Gamma(\omega) = \int_0^\infty \mathrm{d}t\:\alpha( t)\mathrm{e}^{\ii\omega t} = \frac{1}{2}\gamma(\omega)+ \ii S(\omega)
\end{equation}
as it often appears in perturbative master equations. The real part of the above expression is the bath power spectral density, given as
\begin{equation}
    \gamma(\omega)=\frac{2\pi J(|\omega|)}{|1-\exp(-\beta \omega)|}.
\end{equation}
This function determines the decay rates in the usual second-order perturbative master equations. At small $\beta$ (high temperature) and large cutoff frequency, one recovers an effectively flat spectrum
\begin{equation}
    \gamma(\omega)\approx 2\pi J(|\omega|)/|\beta \omega| \approx 2\pi\frac{\alpha}{\beta}  
\end{equation}
leading to Markovian dissipation.

\section{Perturbative Floquet master equations} \label{sec:master_equations}

\subsection{Time-independent case} \label{time_ind}
We first recapitulate the standard steps for deriving the usual Redfield and quantum optical master equation for the case of no driving $H_\mathrm{drive}=0$, as can be found for instance in Ref.~\cite{hartmann_accuracy_2020}. We start by switching to the interaction picture with respect to $H_\mathrm{sys}=H_0$ and $H_\mathrm{env}$. Then, the time evolution of the total state is given by the von Neumann equation
\begin{equation}
    \dot{\tilde \rho}_\mathrm{tot}(t) = -\ii [\tilde H_\mathrm{int}(t), \tilde\rho_\mathrm{tot}(t)]
\end{equation}
where the tilde denotes the interaction picture. To derive a second-order master equation, one inserts the Dyson series for $\tilde{\rho}_\mathrm{tot}(t)$ and traces over the bath to obtain
\begin{equation}\label{eq:Dyson_Redfield}
    \dot{\tilde \rho}(t) = - \tr_\mathrm{env} \int_0^t \mathrm{d}s\; [\tilde H_\mathrm{int}(t), [\tilde H_\mathrm{int}(s), \tilde \rho_\mathrm{tot}(s)]].
\end{equation}
The right-hand side of this equation is already of second order in the coupling strength, so that within a second order perturbation theory we can replace $\tilde\rho_\mathrm{tot}(s)\approx \tilde\rho(t)\otimes \rho_\mathrm{env}$ under the integral. 
Inserting the Gaussian bath model \eqref{eq:H_total}, one obtains
\begin{equation} \label{eq:lindblad_start}
    \dot{\tilde \rho}(t) = - \int_0^t \mathrm{d}s\: \{ \alpha(t-s) [\tilde L(t), \tilde L(s) \tilde \rho(t)]+\mathrm{H.c.}  \}.
\end{equation}
Switching back to the Schr\"odinger picture and substituting $\tau = t-s$ in the integral gives
\begin{equation} \label{eq:schrodinger_start}
\begin{split}
    \dot\rho(t) = &-\ii [H_0, \rho(t)]\\
                  &+\int_0^t \mathrm{d}\tau\: \{ \alpha(\tau) [\tilde L(-\tau)\rho(t), L] + \mathrm{H.c.} \}.
\end{split}
\end{equation}
We assume that the bath correlation function decays sufficiently fast so that the limit of the integral can be extended to $t\rightarrow\infty$.
Then the expression can be further simplified by expanding the coupling operator $L$ in the eigenbasis of $H_0$, which gives
\begin{equation} \label{eigen_decomp}
    \tilde L(t) = \mathrm{e}^{\ii H_0 t} L \mathrm{e}^{-\ii H_0 t} = \sum_{\omega = \epsilon'-\epsilon} \mathrm{e}^{-\ii \omega t} \underbrace{\ketbra{\epsilon}{\epsilon} L \ketbra{\epsilon'}{\epsilon'}}_{L_\omega},
\end{equation}
where the sum runs over all pairs $(\epsilon,\epsilon')$ of eigenvalues of $H_0$ with transition frequencies $\omega=\epsilon'-\epsilon$. Inserting this into Eq.~\eqref{eq:schrodinger_start}, we obtain the Redfield master equation with time-independent coefficients
\begin{equation} \label{eq:redfield}
\begin{split}
      &\dot \rho(t) = -\ii[H_0, \rho(t)] + \sum_{\omega} \Gamma(\omega)[L_{\omega} \rho(t), L] + \mathrm{H.c.}.
\end{split}
\end{equation}
It is well-known that this equation does not describe completely positive dynamics and further approximations are necessary to achieve a GKSL form. One first uses expression \eqref{eq:lindblad_start} and applies the expansion \eqref{eigen_decomp} leading to
\begin{equation}
    \dot{\tilde\rho}(t) = \sum_{\omega, \omega'} \mathrm{e}^{-\ii (\omega-\omega')t} \Gamma(\omega') [L_{\omega'} \tilde\rho(t), L_\omega^\dagger] + \mathrm{H.c.}\:.
\end{equation}
Next, we perform the so-called secular approximation. If all transition frequencies are much larger than the relaxation timescale $|\omega - \omega'| \gg \gamma$ the rapidly oscillating off-diagonal terms in the sum average to zero and can be neglected. Keeping only terms where \mbox{$\omega=\omega'$} and switching back to the Schr\"odinger picture, yields the quantum-optical master equation in the GKSL form
\begin{equation}
\begin{split}\label{eq:Lindblad_undriven}
    \dot{\rho}(t) = &-\ii[H_0 + \sum_{\omega} S(\omega) L^{\dagger}_{\omega}  L_{\omega}, \rho(t) ] \\
    &+ \sum_{\omega}  \gamma(\omega)\left(L_{\omega} \rho(t)L^{\dagger}_{\omega} -\frac{1}{2}\{L^{\dagger}_{\omega}L_{\omega},\rho(t)\}\right),
\end{split}
\end{equation}
including a Lamb-shift contribution to the Hamiltonian.

\subsection{Floquet theorem and Magnus expansion} \label{sec:magnus}

When a driving of the system is included the derivation of master equations becomes more intricate. We consider here specifically the case of periodic driving, which is naturally treated within the framework of Floquet theory. In the Floquet theory of closed quantum systems the Hamiltonian is time-dependent and periodic, i.e. $H(t) = H(t+T)$ which allows one to make use of the Floquet theorem \cite{bukov_universal_2015, mori_floquet_2023}. It states that the time evolution operator can be written as
\begin{equation} \label{eq:floquet_theorem}
    U(t,s) = \mathrm{e}^{-\ii K_\mathrm{F}(t)} \mathrm{e}^{-\ii H_\mathrm{F}\cdot (t-s)} \mathrm{e}^{\ii K_\mathrm{F}(s)}
\end{equation}
where the time-independent Floquet Hamiltonian $H_\mathrm{F}$ governs the slow (stroboscopic) dynamics between the periods and the periodic time-dependent kick operator $K_\mathrm{F}(t)=K_\mathrm{F}(t+T)$ generates the fast dynamics within one period, known as the micromotion. We choose here a particular representation (Floquet gauge) such that $K_\mathrm{F}(0)=0$, although other common choices exist \cite{bukov_universal_2015}.

A description of the problem via the Floquet-Hamiltonian $H_\mathrm{F}$ makes it possible to describe the dynamics of the driven system within a time-independent formalism. An analytical expression for the full Floquet Hamiltonian can only be derived for a few solvable example problems \cite{bukov_universal_2015}. For generic systems one has to resort to numerical integration or perturbative inverse frequency expansions. One possibility to compute $H_\mathrm{F}$ and $K_\mathrm{F}(t)$ in the high-frequency limit is the so-called Magnus expansion. It defines an expansion
\begin{equation}
    H_\mathrm{F} = \sum_{n=0}^\infty H^{(n)}_\mathrm{F}, \quad K_\mathrm{F}(t) = \sum_{n=0}^\infty K^{(n)}_\mathrm{F}(t)
\end{equation}
where the $n$'th term in the sum is of $n$'th order in the inverse driving frequency. 

The Magnus expansion can be used directly on the level of the full Hamiltonian in order to derive a set of Floquet master equations that we call Magnus-Redfield and Magnus-Lindblad equations. For this we utilize the Floquet theorem \eqref{eq:floquet_theorem} for the total Hamiltonian. In particular, we consider the full unitary $U_\mathrm{tot}(t,0) = \mathrm{e}^{-\ii K_\mathrm{tot,F}(t)} \mathrm{e}^{-\ii H_\mathrm{tot,F} t}$. We move to the Floquet reference frame via the transformation
\begin{equation} \label{eq:final_magnus}
    \bar\rho_\mathrm{tot}(t) = \mathrm{e}^{\ii K_\mathrm{tot,F}(t)} \rho_\mathrm{tot}(t) \mathrm{e}^{-\ii K_\mathrm{tot,F}(t)}.
\end{equation}
In this frame the dynamics of the total state is generated by the time-independent Floquet Hamiltonian
\begin{equation}
    \dot{\bar\rho}_\mathrm{tot}(t) = -\ii [H_\mathrm{tot,F}, \bar\rho_\mathrm{tot}(t)].
\end{equation}
The Floquet Hamiltonian can again be written in the generic form \begin{equation}
    H_\mathrm{tot,F}=H'_0\otimes\id_\mathrm{env}+H'_\mathrm{int}+\id_\mathrm{sys}\otimes H'_\mathrm{env}.
\end{equation}
We focus here on the first nontrivial order of the expansion in Hamiltonian \eqref{eq:H_total}. Then we can perform a standard derivation of the Redfield and GKSL master equations as in the time-independent case (Sec.~\ref{time_ind}) and then transform the resulting density operator back to the lab frame. Note that this back-transformation is possible when the kick operator $K_\mathrm{F}$ acts only on the system subspace, which is the case for the lowest nontrivial orders. Note, however, that in general this approach can no longer be used directly at arbitrary order of the Magnus expansion.

At first orders in the Magnus expansion \cite{bukov_universal_2015} one obtains a renormalized model with modified system Hamiltonian and coupling operator 
\begin{equation} \label{eq:magnus_h0}
\begin{split}
    &H_0'=H_0\\
    &+\frac{1}{\omega_d}\sum_{l=1}^\infty \frac{1}{l} \Big([H_l, H_{-l}] +[H_{-l}, H_0] - [H_l, H_0] \Big)
\end{split}
\end{equation}
\begin{equation} \label{eq:magnus_L}
    L'=L+\frac{1}{\omega_d}\sum_{l=1}^\infty \frac{1}{l} \Big([H_{-l}, L] - [H_l, L]\Big)
\end{equation}
but the bath remains identical $H_\mathrm{env}' = H_\mathrm{env}$. In the first-order Magnus expansion the kick operator acts only on the system subspace and takes the form
\begin{equation}\label{eq:Kf_magnus}
    K_\mathrm{tot,F}(t)=K_\mathrm{F}(t)= \frac{1}{\ii \omega_d} \sum_{l\neq 0} H_l \frac{\mathrm{e}^{\ii l\omega_d t}-1}{l}.
\end{equation}
Since the stroboscopic dynamics via $H_\mathrm{tot}'$ is time-independent, we can follow the same steps as in Section \ref{time_ind} to arrive at the \textit{Magnus-Redfield equation}
\begin{equation} \label{eq:redfield_magnus}
\begin{split}
      &\dot {\bar\rho}(t) = \\ & -\ii[H'_0, \tilde\rho(t)] + \sum_{\omega'} (\Gamma(\omega')[L_{\omega'}'\bar\rho(t), L'] + \mathrm{H.c.})
\end{split}
\end{equation}
where $\omega'$ are the transition frequencies of $H'_{0}$.

Likewise, by additionally performing the rotating wave approximation, one can obtain a master equation in GKSL form
\begin{equation} \label{eq:lindblad_magnus}
\begin{split}    
    &\dot{\bar{\rho}}(t) =\\ &-\ii[H_0' + \sum_{\omega'} S(\omega') (L'_{\omega'})^{\dagger} L_{\omega'}', \bar\rho(t) ] \\
    &+ \sum_{\omega'}  \gamma(\omega')\left(L'_{\omega'} \bar\rho(t)(L'_{\omega'})^{\dagger} -\frac{1}{2}\{(L'_{\omega'})^{\dagger}L'_{\omega'},\bar\rho(t)\}\right).
\end{split}
\end{equation}
We will refer to the above as \textit{Magnus-Lindblad equation}. Note that these equations describe only the stroboscopic dynamics. In order to obtain the ``micromotion'' within a period one has to apply the kick-operator $K_\mathrm{tot,F}$. If this operator acts only on the system, as in the first order expansion \eqref{eq:Kf_magnus}, this step can be done entirely on the level of the system density matrix
\begin{equation}
    \rho(t)=\eul^{-\ii K_\mathrm{F}(t)}\bar\rho(t)\eul^{\ii K_\mathrm{F}(t)}.
\end{equation}

\subsection{Weak driving}
Another commonly used approach to incorporate time-dependence into a master equation is to simply replace the system Hamiltonian in the unitary part of the quantum optical master equation while keeping the dissipator from the undriven master equation unchanged \cite{schnellThereFloquetLindbladian2020,schnellHighfrequencyExpansionsTimeperiodic2021a,kolisnyk_floquet_2024, ritter_autonomous_2025, bernazzani_universal_2025}. This can be justified under the assumption that both the system-bath coupling and the driving strength are weak.

The derivation starts similarly as for the time-independent case. We switch to the interaction picture with respect to $H_0$ and $H_\mathrm{bath}$. The total Hamiltonian is then given by $\tilde H_\mathrm{tot}(t) = \tilde H_\mathrm{drive}(t) + \tilde H_\mathrm{int}(t)$ and the von Neumann equation reads
\begin{equation}
    \dot{\tilde \rho}_\mathrm{tot}(t) = -\ii [\tilde H_\mathrm{drive}(t), \tilde \rho_\mathrm{tot}(t)] -\ii [\tilde H_\mathrm{int}(t), \tilde \rho_\mathrm{tot}(t)].
\end{equation}
Leaving the first commutator as is and inserting the formal solution of $\tilde\rho(t)$ in the second, we obtain
\begin{equation}
\begin{split}
    \dot{\tilde \rho}(t) = &-\ii [\tilde H_\mathrm{drive}(t), \tilde \rho(t)]\\
    &- \tr_\mathrm{env} \int_0^t \mathrm{d}s\; [\tilde H_\mathrm{int}(t), [\tilde H_\mathrm{tot}(s), \tilde \rho(s)\otimes \rho_\mathrm{env}]]
\end{split}
\end{equation}
where we replaced $\tilde\rho_\mathrm{tot}(s)$ by $\tilde \rho(s)\otimes \rho_\mathrm{env}$ assuming that the interaction and the driving are at the same perturbative order $\varepsilon_d/\omega\propto \alpha\ll1$.

Using that $\tr_\mathrm{env} \tilde H_\mathrm{int}(t)\rho_\mathrm{env}=0$, we obtain the same dissipator as in the undriven Redfield equation \eqref{eq:Dyson_Redfield}. Following the same derivation, we obtain the GKSL master equation
\begin{equation}\label{eq:lindblad_weak_drive}
    \begin{split}    
    \dot{\rho}(t) = &-\ii[H_\mathrm{sys}(t) + \sum_{\omega} S(\omega) L^{\dagger}_{\omega}  L_{\omega}, \rho(t) ] \\
    &+ \sum_{\omega}  \gamma(\omega)\left(L_{\omega} \rho(t)L^{\dagger}_{\omega} -\frac{1}{2}\{L^{\dagger}_{\omega}L_{\omega},\rho(t)\}\right).
\end{split}
\end{equation}
Note that Lindblad jump operators $L_\omega$ are the original Lindblad operators of the undriven master equation \eqref{eq:Lindblad_undriven}, and the driving is included directly in the unitary part of the evolution. We will refer to the above as \textit{weakly driven Lindblad equation}.

\subsection{Floquet master equations}
The previous approaches included the driving of the system perturbatively. It is also possible to derive a Floquet-Redfield master equation that is non-perturbative in the driving. We repeat here the derivation presented in Ref.~\cite{mori_floquet_2023}. 

We first follow the steps presented in Section \ref{time_ind} up to Eq.~\eqref{eq:lindblad_start}. The next step, switching to Sch\"odinger picture, now requires using the transformation generated by the time-dependent Hamiltonian $H_\mathrm{sys}(t) = H_0+H_\mathrm{drive}(t)$ including the drive. Defining the corresponding unitary
\begin{equation}
    U_\mathrm{sys}(t,s) = \mathcal{T}\mathrm{e}^{-\ii \int_s^t \mathrm{d}\tau \: H_\mathrm{sys}(\tau)}.
\end{equation}
we obtain an expression similar to Eq.~\eqref{eq:schrodinger_start}
\begin{equation} \label{eq:redfield_floquet_schroedinger}
\begin{split}
    \dot\rho(t) = &-\ii [H_\mathrm{sys}(t), \rho(t)]\\
                  &+ \int_0^\infty \mathrm{d}\tau\:  \alpha(\tau) [L(t, t-\tau)\rho(t), L] + \mathrm{H.c.},
\end{split}
\end{equation}
where $L(t,s)$ is the coupling operator in the interaction picture
\begin{equation}
    L(t, t-\tau) = U_\mathrm{sys}(t, t-\tau) \:L\: U^\dagger_\mathrm{sys}(t, t-\tau).
\end{equation}
We now switch to a Floquet reference frame. In contrast to the global Floquet approach presented in Section \ref{sec:magnus}, we define the Floquet Hamiltonian $H_\mathrm{F}$ and the kick operator $K_\mathrm{F}(t)$ now for the reduced system only, i.e.~considering only $H_\mathrm{sys}(t)$. We set $R(t) = \int_0^\infty \mathrm{d}\tau\: \alpha(\tau) L(t, t-\tau)$ and define the transformation to the Floquet reference frame (indicated by an overline)
\begin{equation}
\begin{split}
    \bar\rho(t) &= \mathrm{e}^{\ii K_\mathrm{F}(t)} \rho(t) \mathrm{e}^{-\ii K_\mathrm{F}(t)}\\
    \bar L(t) &= \mathrm{e}^{\ii K_\mathrm{F}(t)} L \mathrm{e}^{-\ii K_\mathrm{F}(t)} = \sum_{n = -\infty}^{+\infty} \bar L_n \mathrm{e}^{\ii n \omega_d t}\\
    \bar R(t) &= \mathrm{e}^{\ii K_\mathrm{F}(t)} R(t) \mathrm{e}^{-\ii K_\mathrm{F}(t)} = \sum_{n = -\infty}^{+\infty} \bar R_n \mathrm{e}^{\ii n \omega_d t}.
\end{split}
\end{equation}
The expansion in a Fourier series is justified since all operators are periodic with the frequency of the drive $\omega_d$ due to the periodicity of $K_\mathrm{F}(t)$. Applying this transformation to Eq.~\eqref{eq:redfield_floquet_schroedinger} we obtain the \textit{Floquet-Redfield equation} \cite{mori_floquet_2023}
\begin{equation} \label{eq:time-dependent-floquet-redfield}
\begin{split}
    &\dot{\bar\rho}(t) = -\ii[H_\mathrm{F}, \bar\rho(t)]\\
    & + \sum_{n, m = -\infty}^{+\infty} \mathrm{e}^{\ii(n+m)\omega_d t} \left\{[\bar R_n \bar\rho(t), \bar L_m] + [\bar L_m, \bar\rho(t) \bar R^\dagger_{-n}] \right\}
\end{split}
\end{equation}
where 
\begin{equation} \label{eq:R_def}
    \bar R_n = \int_0^{\infty} \mathrm{d}\tau\: \alpha(\tau) \mathrm{e}^{-\ii H_\mathrm{F} \tau} \bar L_n \mathrm{e}^{\ii H_\mathrm{F} \tau} \mathrm{e}^{-\ii n\omega_d \tau}.
\end{equation}
Note that, in contrast to the Redfield master equation for undriven systems, the total dissipator of the Floquet-Redfield equation is now a sum of many time-dependent components.

In order to obtain a time-independent generator one neglects the time-dependent contributions in \eqref{eq:time-dependent-floquet-redfield} using a rotating-wave argument. If $\omega_d$ is sufficiently large, terms in the master equation where $n\neq-m$ are rapidly oscillating and can therefore be discarded. Keeping only the terms with $n=-m$ results in a \textit{time-independent Floquet-Redfield equation}
\begin{equation} \label{eq:floquet-redfield}
\begin{split}
    \dot{\bar\rho}(t) &= -\ii[H_\mathrm{F}, \bar\rho(t)]\\
    & + \sum_{n = -\infty}^{+\infty} \left\{[\bar R_{-n} \bar\rho(t), \bar L_n] + [\bar L_n, \bar\rho(t) \bar R^\dagger_{n}] \right\}.
\end{split}
\end{equation}
It is also possible to enforce a GKSL form of the Floquet-Redfield equation. We transform the above equation into the interaction picture with respect to the Floquet Hamiltonian. Expanding operators in the eigenbasis of $H_\mathrm{F}$ yields (similar to Eq.~\eqref{eigen_decomp})
\begin{equation} \label{eq:eigen_decomp_LR}
\begin{split}
    \tilde{\bar L}_n(t) &= \mathrm{e}^{\ii H_\mathrm{F}t} \bar L_n \mathrm{e}^{-\ii H_\mathrm{F}t} \\
    &= \sum_{\omega_\mathrm{F}=\epsilon'_\mathrm{F}-\epsilon_\mathrm{F}}\mathrm{e}^{-\ii\omega_\mathrm{F}t} \underbrace{\ketbra{\epsilon_\mathrm{F}}{\epsilon_\mathrm{F}} \bar L_n \ketbra{\epsilon'_\mathrm{F}}{\epsilon'_\mathrm{F}}}_{\bar L_n[\omega_\mathrm{F}]}\\
    \tilde{\bar R}_n(t) &= \mathrm{e}^{\ii H_\mathrm{F}t} \bar R_n \mathrm{e}^{-\ii H_\mathrm{F}t} \\
    &= \sum_{\omega_\mathrm{F}=\epsilon'_\mathrm{F}-\epsilon_\mathrm{F}}\mathrm{e}^{-\ii\omega_\mathrm{F}t} \underbrace{\ketbra{\epsilon_\mathrm{F}}{\epsilon_\mathrm{F}} \bar R_n \ketbra{\epsilon'_\mathrm{F}}{\epsilon'_\mathrm{F}}}_{\bar R_n[\omega_\mathrm{F}]}
\end{split}
\end{equation}
where tilde denotes the interaction picture and $\varepsilon_F$ are the eigenenergies of $H_F$. The master equation then takes the form
\begin{equation}
\begin{split}
    \dot{\tilde{\bar\rho}} =\sum_{n=-\infty}^{+\infty}\sum_{\omega_F,\omega_F'}  \mathrm{e}^{-\ii (\omega_\mathrm{F}-\omega'_\mathrm{F})t} \Big\{ &[\bar R_{-n}[-\omega'_\mathrm{F}] \tilde{\bar\rho}(t), \bar L_n[\omega_\mathrm{F}]] \\
    + &[\bar L_n[\omega_\mathrm{F}], \tilde{\bar\rho}(t) \bar R^\dagger_n[\omega'_\mathrm{F}]] \Big\}.
\end{split}
\end{equation}
Applying the secular approximation $ |\omega_\mathrm{F} - \omega'_\mathrm{F}| \gg \gamma$, only terms $\omega_\mathrm{F} = \omega'_\mathrm{F}$ remain. Finally, by combining Eq.~\eqref{eq:R_def} and Eq.~\eqref{eq:eigen_decomp_LR} 
\begin{equation}
\begin{split}
    \bar R_n[\omega_\mathrm{F}] &= \bar L_n[\omega_\mathrm{F}]\int_0^\infty \mathrm{d}\tau\:\alpha(\tau)\mathrm{e}^{\ii(\omega_\mathrm{F}-n\omega_d)t}\\
    &= \bar L_n[\omega_\mathrm{F}] \left(\frac{1}{2}\gamma(\omega_F-n\omega_d) + \ii S(\omega_F-n\omega_d)\right)
\end{split}
\end{equation}
we arrive at a Floquet master equation in the GKSL form
\begin{equation}
\begin{split}\label{eq:lindblad_floquet}
    &\dot{\bar\rho}(t) =\\& -\ii[H_\mathrm{F} + H_\mathrm{LS}, \bar\rho(t)]
    + \sum_{n=-\infty}^{+\infty} \sum_{\omega_\mathrm{F}} \gamma(\omega_\mathrm{F}-n\omega_d) \\
    &\times \left(\bar L_n[\omega_\mathrm{F}] \bar\rho(t) \bar L^\dagger_n[\omega_\mathrm{F}] - \frac{1}{2}\{ \bar L^\dagger_n[\omega_\mathrm{F}] \bar L_n[\omega_\mathrm{F}], \bar\rho(t)\}\right)
\end{split}
\end{equation}
where 
\begin{equation}
    H_\mathrm{LS} = \sum_{n=-\infty}^{+\infty} \sum_{\omega_\mathrm{F}} S(\omega_\mathrm{F}-n\omega) \bar L^\dagger_n[\omega_\mathrm{F}] \bar L_n [\omega_\mathrm{F}].
\end{equation}
We will refer to the above as \textit{Floquet-Lindblad equation}. 

Note, that the dissipators of the (time-independent) Floquet-Redfield and Floquet-Lindblad equations consist of many Fourier components. In real calculations, these sums can be truncated numerically because the norm of the operators $\bar L_n$ or $\bar{R}_n$ vanishes at large $n$.

An overview of the different master equations and the underlying approximations is displayed in Tab.~\ref{tab:overview}.

\begin{table}[]
    \centering
\begin{tabular}{|c|c|c|c|c|}
    \hline
     Master equation & \quad Eq.~ \quad & \, slow  \,&  strong  & \, near \, \\
    & Ref. & \,drive\, &\, drive \,& deg. \\ \hline
     Floquet-Redfield & \eqref{eq:time-dependent-floquet-redfield} & {\color{blue}\ding{51}} & {\color{blue}\ding{51}} & {\color{blue}\ding{51}}  \\
     time-ind.~Floquet-Redfield & \eqref{eq:floquet-redfield} & {\color{red}\ding{55}} & {\color{blue}\ding{51}} & {\color{blue}\ding{51}}  \\

     Magnus-Redfield   & \eqref{eq:redfield_magnus} & {\color{red}\ding{55}} & {\color{red}\ding{55}} & {\color{blue}\ding{51}} \\
     Floquet-Lindblad & \eqref{eq:lindblad_floquet}& {\color{red}\ding{55}}  & \,\,\,{\color{blue}\ding{51}}$\,^*$ & {\color{red}\ding{55}} \\
     Magnus-Lindblad  & \eqref{eq:lindblad_magnus}& {\color{red}\ding{55}} & {\color{red}\ding{55}} & {\color{red}\ding{55}} \\
     weakly driven Lindblad & \eqref{eq:lindblad_weak_drive}& {\color{blue}\ding{51}}   & {\color{red}\ding{55}} & {\color{red}\ding{55}} \\ \hline
\end{tabular}
    \caption{Overview of the different perturbative master equations considered in this paper and their regime of applicability. Master equations that require the secular approximation become unreliable when the bare energy spectrum of the system is nearly-degenerate (near deg.). All master equations require weak coupling to the bath. $\!\,^*$In the Floquet-Lindblad equation, even though there is no explicit assumption of weak driving, the driving strength (as well as the driving frequency) indirectly affects the validity of the secular approximation.}
    \label{tab:overview}
\end{table}

\section{Error measure and Reference solution} \label{sec:error_convergence}

\subsection{Error measure for dynamical maps}

In order to rigorously compare the performance of master equations we introduce an error measure that bounds the maximum error on any simulation result irrespective of the initial condition. 
Consider a reduced state $\rho(t)$ calculated using any of the master equations and an exact reference state $\rho_\mathrm{ref}(t)$. Then the error of an arbitrary observable $O$, can be bounded as follows \cite{watrousTheoryQuantumInformation2018}
\begin{equation}
\begin{split}
    &|\delta \langle O(t) \rangle| = \\
    &|\tr[(\rho_\mathrm{ref}(t)-\rho(t))O]| \leq ||O||_\infty\;||\rho_\mathrm{ref}(t)-\rho(t)||_1
\end{split}
\end{equation}
where $||O||_\infty$ is the largest singular value of $O$ and $||\cdot||_1$ is the trace norm.
Quantum master equations predict the dynamical map, that is the map from the initial to the final system state. We write this linear map as a superoperator $\Phi(t)$ acting on a density matrix as $\rho(t)=\Phi(t)\rho(0)$. Thus we can write
\begin{equation}
    ||\rho_\mathrm{ref}(t)-\rho(t)||_1=||[\Phi_\mathrm{ref}(t)-\Phi(t)]\rho(0)||_1.
\end{equation}
Note that this is tightly bounded by
\begin{equation}
    ||(\Phi(t)-\Phi_\mathrm{ref}(t))\rho(0)||_1\leq ||\Phi(t)-\Phi_\mathrm{ref}(t)||_{1\rightarrow 1}
\end{equation}
where $||\cdot||_{1\rightarrow 1}$ denotes the operator norm induced by the trace norm. This induced operator norm is difficult to compute numerically. A simple upper bound can be given by the maximum singular value $\sigma_\mathrm{max}(\cdot)$ of the superoperator \cite{watrousTheoryQuantumInformation2018}
\begin{equation}
    ||\Phi(t)-\Phi_\mathrm{ref}(t)||_{1\rightarrow 1}\leq\sqrt{d}\,\sigma_\mathrm{max}(\Phi(t)-\Phi_\mathrm{ref}(t)),
\end{equation}
where $d$ is the dimension of the system Hilbert space ($d=4$ for two spins). Thus we obtain an efficiently computable upper bound for the error of any observable through
\begin{equation}
    \frac{|\delta \langle O(t) \rangle| }{||O||_\infty}\leq \varepsilon(t)=\sqrt{d}\,\sigma_\mathrm{max}(\Phi_\mathrm{ref}(t) - \Phi(t))
\end{equation}
For a full evolution we consider as our error measure the maximum value of $\varepsilon(t)$ throughout the evolution 
\begin{equation}\label{eq:error_bound}
    \varepsilon = \max_{t} \varepsilon(t),
\end{equation}
such that the error of any observable at any time is bounded by $\varepsilon$. In our simulations we found that using different error measures leads to qualitatively similar results, but these may not provide such a tight bound on arbitrary observables.

\subsection{Exact reference solution}
The exact dynamics of the systems can be obtained from numerical techniques for non-Markovian open quantum systems \cite{devegaDynamicsNonMarkovianOpen2017,Hartmann2017hops,tanimuraNumericallyExactApproach2020, Hartmann2021hops,ortega-tabernerUnifyingMethodsOptimal2024,chinAlgorithmsSoftwareOpen2025,keelingProcessTensorApproaches2025}. These methods are accurate and well-established and can provide a reference for benchmarking of master equations. 

Here we utilize an exact method based on tensor network influence functionals, building upon the time-evolving matrix product operator (TEMPO) approach \cite{strathearnEfficientNonMarkovianQuantum2018}. We use specifically the highly efficient uniform TEMPO algorithm from Ref.~\cite{linkOpenQuantumSystem2024a} which allows for accurate simulations to arbitrary long times \cite{kahlertSimulatingLandauZener2024, garbelliniUniformProcessTensor2026a}. The method can handle local time-dependence \cite{fuxEfficientExplorationHamiltonian2021,kahlertSimulatingLandauZener2024} and can also be directly combined with Floquet theory, as described in Ref.~\cite{mickiewicz_exact_2025}. 

The parameters relevant for convergence of the method are the evolution time step $\delta t$ (Trotter step) and the so-called bond dimension $\chi$, representing the dimension of the memory space used to model exact non-Markovian response of the full environment. In the weak-coupling regime the method converges very quickly towards the exact solution. 

\begin{figure}
    \centering
    \includegraphics[scale=1]{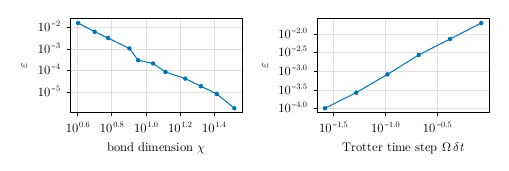}
    \caption{Convergence of the dynamical map for the driven two-spin boson mode obtained with uniTEMPO with respect to the bond dimension (left panel) and Trotter time step (right panel). As a reference we used high-accuracy calculations with large bond dimension and small Trotter step. Model parameters are $\epsilon_d = 5\Omega$, $\omega_d = 10\Omega$, $\omega_A=\omega_B=\Omega$, $\alpha=2.5\times 10^{-4}$, $\omega_c=20\Omega$, $\beta=0.1/\Omega$.}\label{fig:convergence}
\end{figure}

In Fig. \ref{fig:convergence} we show the convergence of uniTEMPO with respect to the reference solution, quantified with the error measure $\varepsilon$ from Eq.~\eqref{eq:error_bound}. The reference $\Phi_\mathrm{ref}$ was computed using a large bond dimension (left panel) or a small Trotter step (right panel), whereas $\Phi$ was calculated with varying parameters $\chi$ and $\delta t$. In all cases, the dynamical maps were computed up to time $t=1000/\Omega$. We performed the simulations for strong and fast driving in the weak-coupling and Markovian regime. We find rapid convergence with respect to both parameters.

While simulations are inexpensive for the two-spin-boson model considered in this article, these methods become computationally demanding for larger systems \cite{keelingProcessTensorApproaches2025,linkTensorNetworkInfluence2026}. Moreover, they often make it difficult to identify the physical mechanisms that are introduced through weak dissipation. For these reasons, master equations remain an important theoretical and numerical tool for weakly coupled open systems.

\section{Benchmarks of the Floquet master equations} \label{sec:results}

Detailed assessments of the performance of master equations with respect to the system-bath coupling and the bath structure for the case of undriven dynamics can be found in Refs.~\cite{hartmann_accuracy_2020, tellobreuerBenchmarkingQuantumMaster2024}. In this article, we compare the performance of master equations specifically with respect to the periodic time-dependent driving in a Floquet setting. We therefore choose parameters for the dissipation in such a way that all master equations perform well in the absence of driving. This generally requires weak coupling, the absence of near-degeneracies in the spectrum of $H_0$, as well as high temperatures.

In the following sections we calculate the error measure \eqref{eq:error_bound} for each master equation as a function of the driving parameters: the frequency $\omega_d$ and the amplitude $\epsilon_d$. For each set of driving parameters, we compute the dynamical maps predicted by the master equations and uniTEMPO up to time $t=1000/\Omega$. Based on our observations of the dynamics of various observables, we adopt $\varepsilon=0.2$ as the maximum acceptable error. Results with error exceeding this threshold are not discussed. 

We note that, for the displayed examples, using the full time-dependent Floquet-Redfield equation \eqref{eq:time-dependent-floquet-redfield}, while numerically relatively demanding, consistently lead to errors below $\varepsilon=0.025$, independent of the driving strength and driving frequency. This is expected because the equation relies solely on the Born approximation and no assumption on the driving strength is necessary. Therefore, we do not display the error for this equation in the following.

\subsection{Example dynamics}
\begin{figure}
    \centering
    \includegraphics[scale=1.]{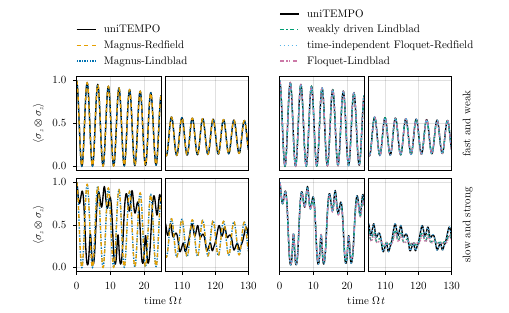}
    \caption{Expectation value of $\sigma_z\otimes\sigma_z $ for an example dynamics of the driven two-spin boson model with initial state $\rho(0) = \ketbra{11}{11}$. Top row: fast and weak driving regime ($\Omega/\omega_d=0.1,\:\epsilon_d = 0.1\Omega$) where our error measure is small for all master equations $\varepsilon < 0.05$. Bottom row: slow and strong driving regime ($\Omega/\omega_d=1.3,\:\epsilon_d = 4\Omega$) where our error measure of the master equations is large $\varepsilon > 0.2$, with an exception of the time-independent Floquet-Redfield for which $\varepsilon < 0.05$. Remaining simulation parameters: $\omega_A=\omega_B=\Omega$, $\alpha=2.5\times 10^{-4}$, $\omega_c=20\Omega$, $\beta=0.1/\Omega$.
    }\label{fig:dynamics}
\end{figure}

As an illustrative example, we first explicitly compute the dynamics of the two-spin-boson model using the master equations and uniTEMPO. In Fig. \ref{fig:dynamics} we show the time evolution of $\langle \sigma_z\otimes\sigma_z \rangle$ in two distinct driving parameter regimes: fast and weak (large $\omega_d$ and small $\epsilon_d$) and slow and strong (small $\omega_d$ and large $\epsilon_d$). In general, we expect all the master equations to be more accurate when the driving is fast and weak, compared to the slow and strong regime. Indeed, we see in Fig. \ref{fig:dynamics} that most of the equations struggle to correctly reproduce the dynamics in the slowly and strongly driven regime. Only the time-independent Floquet-Redfield equation is accurate for both parameter choices, despite the high-frequency requirement (see Table \ref{tab:overview}).

\subsection{Weak driving} \label{sec:weak_drive}

\begin{figure}
    \centering
    \includegraphics[scale=1.]{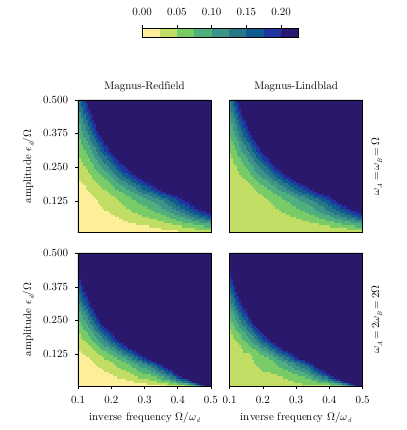}
    \caption{Error measure \eqref{eq:error_bound} for the Magnus-Redfield and Magnus-Lindblad master equations as a function of the driving frequency $\omega_d$ and the driving amplitude $\epsilon_d$ for resonant ($\omega_A=\omega_B=\Omega$, top row) and detuned ($\omega_A=2\omega_B=2\Omega$, bottom row) spins. The validity of the master equations is restricted to the high-frequency and low-amplitude regime (bottom left corner of each plot) as a result of the underlying first-order Magnus expansion. Remaining simulation parameters: $\alpha=2.5\times 10^{-4}$, $\omega_c=20\Omega$, $\beta=0.1/\Omega$.
    }\label{fig:magnus}
\end{figure}

We first consider the weakly driven regime. In Fig.~\ref{fig:magnus} we present the results for the Magnus-Redfield and Magnus-Lindblad master equations. In both cases, the underlying first-order Magnus expansion significantly limits the validity of both master equations with respect to the driving strength and the driving frequency. Generally, as expected, the Magnus-Redfield equation performs better than its Lindblad counterpart. Differences in the case of zero detuning $\omega_A=\omega_B$ compared to $\omega_A=2\omega_B$ are due to differences in the energy scale of the system.

\begin{figure}
    \centering
    \includegraphics[scale=1.]{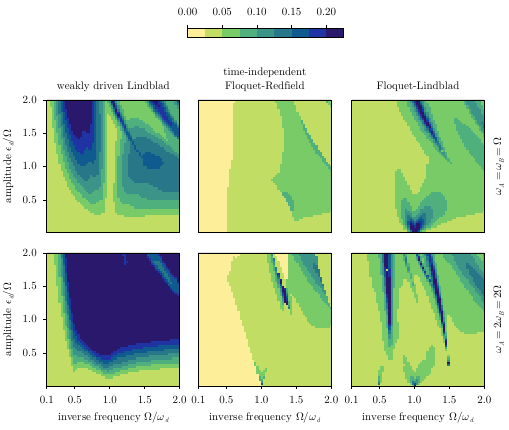}
    \caption{Error measure \eqref{eq:error_bound} for the weakly driven Lindblad, time-independent Floquet-Redfield and Floquet-Lindblad master equations as a function of the driving frequency $\omega_d$ and the driving amplitude $\epsilon_d$ for the resonant ($\omega_A=\omega_B=\Omega$, top row) and detuned ($\omega_A=2\omega_B=2\Omega$, bottom row) spins. The validity differs for each master equation and each regime. We achieve the overall best accuracy for the time-independent Floquet-Redfield equation. Remaining simulation parameters: $\alpha=2.5\times 10^{-4}$, $\omega_c=20\Omega$, $\beta=0.1/\Omega$.
    }\label{fig:floquet_weak}
\end{figure}

We perform a similar parameter scan for the weakly driven Lindblad, time-independent Floquet-Redfield and Floquet-Lindblad master equations. The results are presented in Fig. \ref{fig:floquet_weak}.

For the weakly driven Lindblad master equation we obtain a very good accuracy when the driving amplitude is small, as expected from the perturbative derivation. In the resonant spin case $\omega_A=\omega_B$ we also find high accuracy in the vicinity of the resonance $\omega_d \approx 1\Omega$, even up to moderate values of the driving amplitude. This peak is, however, absent when the spins are detuned.

Among the three master equations, we find that the time-independent Floquet-Redfield equation shows the overall best performance, in agreement with previous studies of time-independent master equations \cite{hartmann_accuracy_2020,tellobreuerBenchmarkingQuantumMaster2024}. For high-frequency driving the error remains low even for strong driving amplitudes.

The Floquet-Lindblad master equation performs poorly in the vicinity of the resonant frequency $\omega_d \approx \Omega$. At the same time it is more accurate in the high-frequency regime, compared to the weakly driven Lindblad equations. A striking feature are isolated regimes where the performance of the Floquet-Lindblad equation drops significantly. These correspond to regions where a small gap emerges in the spectrum of the Floquet Hamiltonian, invalidating the secular approximation.

\subsection{Strong driving}

\begin{figure}
    \centering
    \includegraphics[scale=1.]{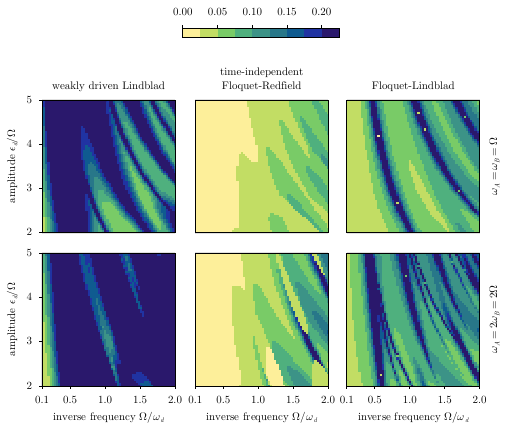}
    \caption{Error measure \eqref{eq:error_bound} for the weakly driven Lindblad, time-independent Floquet-Redfield and Floquet-Lindblad master equations as a function of the driving frequency $\omega_d$ and the driving amplitude $\epsilon_d$. The same parameters as in Fig.~\ref{fig:floquet_weak} but for larger driving amplitudes.}\label{fig:floquet_strong}
\end{figure}

We now consider larger driving amplitudes. In this case we expect most of the master equations to deviate more strongly from the exact solution. We focus here only on the weakly driven Lindblad, time-independent Floquet-Redfield and Floquet-Lindblad master equations, since Magnus-based equations perform very poorly in this regime. Our results are presented in Fig. \ref{fig:floquet_strong}. 

As can be expected, the accuracy of the weakly driven Lindblad equation is very limited at stronger driving due to the explicit perturbative treatment of the time-dependent part of the Hamiltonian.

The time-independent Floquet-Redfield equation remains very accurate in the high-frequency regime, with similar error as in the weakly driven regime in Fig.~\ref{fig:floquet_weak}. When the driving is slow, the performance drops more significantly because off-diagonal components in the full Redfield equation \eqref{eq:floquet-redfield} can no longer be neglected.

Lastly, the accuracy of the Floquet-Lindblad equation is strongly affected by degeneracies in the spectrum of the bare system Floquet-Hamiltonian $H_F$, thus leading to sharp features in our error measure. The error is significantly increased close to degeneracies, while it can remain accurate in non-degenerate regions. This problem is known for the quantum optical master equation \cite{hartmann_accuracy_2020} and can be partially alleviated using so-called coarse grained master equations \cite{benattiEntanglingTwoUnequal2010,  benattiEnvironmentinducedEntanglementRefined2009,schallerPreservationPositivityDynamical2008,majenzCoarseGrainingCan2013}. Generalizing this approach to the Floquet case is an interesting direction for future works.

\subsection{Lower temperature}
 
We further consider a bath at lower temperature where perturbative master equations generally become less accurate. Note that, in order to reach decay rates comparable to the previous examples, the coupling strength $\alpha$ must be increased when the inverse temperature $\beta$ is increased. At lower temperature the bath spectrum deviates more strongly from a flat spectral density, thus generally leading to more non-Markovian dynamics. In the case of zero detuning $\omega_A=\omega_B$ the relaxation dynamics becomes very slow at lower temperature, which causes difficulties in the interpretation of errors at large evolution times. Therefore, we considered only the detuned case $\omega_A=2\omega_B$ for this parameter regime.

\begin{figure}
    \centering
    \includegraphics[scale=1.]{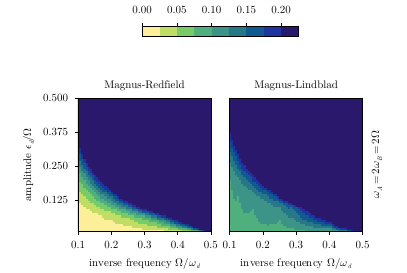}
    \caption{Error measure \eqref{eq:error_bound} for the Magnus-Redfield and Magnus-Lindblad master equations as a function of the driving frequency $\omega_d$ and the driving amplitude $\epsilon_d$, as in Fig.~\ref{fig:magnus} but for lower bath temperature $\beta=1/\Omega$ and $\alpha = 1.25\times10^{-3}$.
    }\label{fig:magnus_lowT}
\end{figure}

We present the results for the Magnus-Redfield and Magnus-Lindblad equations in Fig. \ref{fig:magnus_lowT}. The accuracy of both equations is more limited compared to the high-temperature case in Fig.~\ref{fig:magnus}. In particular, at lower temperatures, the Lindblad equations become inaccurate even in the absence of driving. This is because the error introduced by the secular approximation is amplified when the bath spectrum is more structured.

\begin{figure}
    \centering
    \includegraphics[scale=1.]{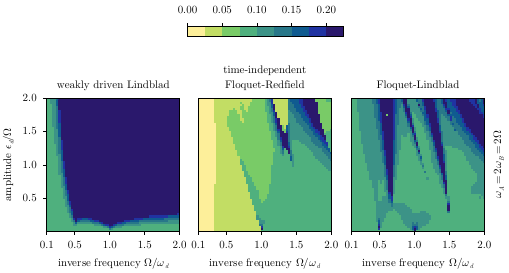}
    \caption{Error measure \eqref{eq:error_bound} for the weakly driven Lindblad, time-independent Floquet-Redfield and Floquet-Lindblad master equations as a function of the driving frequency $\omega_d$ and the driving amplitude $\epsilon_d$, as in Fig.~\ref{fig:floquet_weak} but for lower bath temperature $\beta=1/\Omega$ and $\alpha = 1.25\times10^{-3}$.
    }\label{fig:floquet_weak_lowT}
\end{figure}

\begin{figure}
    \centering
    \includegraphics[scale=1.]{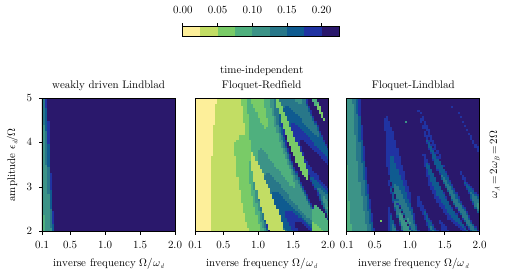}
    \caption{Error measure \eqref{eq:error_bound} for the weakly driven Lindblad, time-independent Floquet-Redfield and Floquet-Lindblad master equations as a function of the driving frequency $\omega_d$ and the driving amplitude $\epsilon_d$, as in Fig.~\ref{fig:floquet_weak} but for lower bath temperature $\beta=1/\Omega$ and $\alpha = 1.25\times10^{-3}$.}\label{fig:floquet_strong_lowT}
\end{figure}

Results for the weakly driven Lindblad, time-independent Floquet-Redfield and Floquet-Lindblad master equations are displayed in Fig. \ref{fig:floquet_weak_lowT} and Fig. \ref{fig:floquet_strong_lowT}. For all three master equations, we find qualitatively similar behavior to the previous high temperature simulations. However, the errors are generally significantly larger compared to the high temperature case.

Overall this analysis shows that, irrespective of the driving, Redfield-level master equations can still perform well at lower temperatures whereas GKSL-type equations become unreliable. The presence of strong or slow driving induces additional errors that restrict the applicability of the various master equations even further.

\section{Conclusions}

In this work, we benchmarked several widely used Floquet master equations against numerically exact uniTEMPO simulations for a periodically driven two-spin-boson model. By comparing the full dynamical maps over long relaxation times, we rigorously quantified how the accuracy of each master equation depends on the driving amplitude and driving frequency. Across all parameter regimes we found that the observed errors closely follow the assumptions entering the derivation of the respective equations. However, estimating a quantitative error without knowledge of the exact solution appears elusive in large parameter regimes.

Our results show that master equations based on perturbative treatments of the drive are reliable only in correspondingly restricted regimes. The Magnus-based approaches are limited to weak and high-frequency driving, while the weakly driven Lindblad equation remains accurate at low-frequencies when the driving is weak. The time-independent Floquet-Redfield equation provides the most robust overall description while remaining computationally inexpensive. It provides accurate results over a broad range of driving amplitudes and frequencies and shows only a weak dependence on the driving frequency. The full time-dependent Floquet-Redfield equation remains accurate throughout the parameter ranges studied here, consistent with the fact that it relies only on the Born approximation and does not require additional assumptions on the drive. However, it is numerically the most demanding master equation as it involves time-dependence and a double-sum over Fourier components in the generator. In general, the secular approximation is a significant source of error in Floquet master equations and should be avoided if possible. In particular, the Floquet-Lindblad equation can be unreliable near resonances and in parameter regions where small quasienergy gaps invalidate the separation of time-scales required for secularization. At lower temperatures GKSL-type equations deteriorate significantly, whereas Redfield-type equations can remain quantitatively reliable.

If a time-independent and numerically efficient description is desired, the time-independent Floquet-Redfield equation appears to offer the best compromise between accuracy and simplicity within the regimes considered here. More broadly, our results show that reproducing driven dissipative dynamics requires careful control not only of weak-coupling errors, but also of the additional approximations introduced by periodic driving. 

Apart from the master equations considered in this work, several other equations have been introduced in the case of undriven systems that eliviate certain problems with the standard Redfield or Lindblad approaches, such as positivity violation \cite{nathanUniversalLindbladEquation2020}, long-time accuracy \cite{beckerCanonicallyConsistentQuantum2022,sartipi2026canonicallyconsistentquantummaster}, or sensitivity to near-degeneracies  \cite{benattiEntanglingTwoUnequal2010,  benattiEnvironmentinducedEntanglementRefined2009,schallerPreservationPositivityDynamical2008,majenzCoarseGrainingCan2013}. These equations have, however, not yet been combined with Floquet theory. It would further be interesting to perform a similar benchmark of master equations in the case of many-body systems where periodic driving and dissipation can create exotic phases of matter \cite{sahaPrethermalDiscreteTime2024,dasEnvironmentassistedDiscreteTime2026,wanckelDissipativeFloquetEngineering2026} therefore making an accurate description of the dynamics desirable. Particularly interesting in a many-body setting are so-called local master equations which avoid an exact diagonalization of the system Hamiltonian \cite{schnellGlobalBecomesLocal2025,shiraishiQuantumMasterEquation2025}. Moreover, a similar analysis as presented in this paper could be performed for the case of fermionic reservoirs commonly studied in transport through quantum dots \cite{harbolaQuantumMasterEquation2006, acciaiQuantumTransportPhenomena2025,sevitz2025quantummasterequationnanoelectromechanical}, with exact reference solutions available via fermionic TEMPO variants \cite{thoennissEfficientMethodQuantum2023,ngRealtimeEvolutionAnderson2023,chenRealtimeImpuritySolver2024,sonnerSemigroupInfluenceMatrices2025}.

\bibliography{bib_paper.bib}

\end{document}